# Construction of a qudit using Schrödinger cat states and generation of hybrid entanglement between a discrete-variable qudit and a continuous-variable qudit

Qi-Ping Su,[1] Tong Liu,[2] Yu Zhang,[3] and Chui-Ping Yang[1, 2,*]

[1]*Department of Physics, Hangzhou Normal University, Hangzhou 311121, China*
[2]*Quantum Information Research Center, Shangrao Normal University, Shangrao 334001, China*
[3]*School of Physics, Nanjing University, Nanjing 210093, China*

We show that a continuous-variable (CV) qudit can be constructed using quasiorthogonal cat states of a bosonic mode, when the phase encoded in each cat state is chosen appropriately. With the constructed CV qudit and the discrete-variable (DV) qudit encoded with Fock states, we propose an approach to generate the hybrid maximally entangled state of a CV qudit and a DV qudit by using two microwave cavities coupled to a superconducting flux qutrit. This proposal relies on the initial preparation of a superposition of Fock states of one cavity and the initial preparation of a cat state of the other cavity. After the initial state of each cavity is prepared, this proposal requires only two basic operations, i.e., the first operation employs the dispersive coupling of both cavities with the qutrit while the second operation uses the dispersive coupling of only one cavity with the qutrit. The entangled state production is deterministic and the operation time decreases as the dimensional size of each qudit increases. In addition, during the entire operation, the coupler qutrit remains in the ground state and thus decoherence from the qutrit is significantly reduced. As an example, we further discuss the experimental feasibility for generating the hybrid maximally entangled state of a DV qutrit and a CV qutrit based on circuit QED. This proposal is universal and can be extended to accomplish the same task, by using two microwave or optical cavities coupled to a natural or artificial three-level atom.

## I. INTRODUCTION AND MOTIVATION

Compared with a qubit, a qudit (i.e., *d*-level or *d*-state quantum system) has a larger Hilbert space and thus can be used to encode more information in quantum information processing and communication. Recently, entangled states of qudits or high-dimensional quantum systems have drawn much attention because they have many practical applications, such as improving the robustness of quantum key distribution [1], simplifying quantum gate implementation [2], increasing the generation rate of quantum random numbers [3], and testing the foundations of quantum mechanics [4]. Stimulated by these prospects, the creation of entangled states of qudits or high-dimensional quantum systems has been extensively investigated in various physical degrees of freedom of light [5–14]. On the other hand, *hybrid* entangled states play an important role in quantum information processing and quantum technology. For examples, hybrid entangled states can be used as quantum channels and intermediate resources for quantum technology, covering quantum information transmission, quantum state manipulation, and storage between different encodings and formats [15–17]. The hybrid entangled states usually involve subsystems which are different in their nature (e.g., photons and matters) or in the degree of freedom (e.g., discrete-variable degree and continuous-variable degree).

The hybrid maximally entangled states of discrete-variable (DV) qudits and continuous-variable (CV) qudits are key resources for building hybrid high-dimensional quantum networks. They can act as quantum channels to transfer high-dimensional quantum states between distant DV qudits and CV qudits and to achieve superdense coding and quantum key distribution. They can also serve as practical interfaces to connect DV-qudit-based quantum processors and CV-qudit-based quantum processors. A DV qudit lives in a high-dimensional Hilbert space formed, for example, by the Fock states with different photon numbers. On the contrary, the encoding for a CV qudit can be realized in the quadrature components of a light field with an inherently infinite-dimensional space. Quantum information carried by a CV qudit can be expressed, for instance, as an arbitrary superposition of coherent states or cat states of light waves. A hybrid combination of DV qudits and CV qudits may provide a huge advantage. Some operations may benefit from DV qudits, while others may be more effective using CV qudits.

Motivated by the above, this work focuses on generation of a hybrid maximally entangled state of a DV qudit and a CV qudit. The CV qudit here is encoded using cat states of a bosonic mode, while the DV qudit here is encoded using Fock states. This proposal is based on circuit QED. As is well known, the new and rapidly growing field of circuit QED, consisting of microwave radiation fields and fixed artificial atoms, offers extremely exciting prospects for solid-state quantum information processing [18–25]. In the following, we first propose to construct a CV qudit using cat states of a bosonic mode, and show that any two of cat states can be made quasiorthogonal to each other when the phase encoded in each cat state is chosen appropriately. With the constructed CV qudit and the DV qudit encoded via Fock states, we then present an efficient method to create a hybrid maximally entangled state of a DV qudit and a CV qudit, by using a circuit-QED setup consisting of two microwave cavities coupled to a superconducting flux qutrit (a three-level artificial atom). This proposal relies on the initial preparation of a superposition of Fock states of one cavity and the initial preparation of a cat state of the other cavity.

*yangcp@hznu.edu.cn

As an example of this proposal, we further discuss the experimental feasibility for generating the hybrid maximally entangled state of a DV qutrit and a CV qutrit (i.e., the case for $d = 3$) based on circuit QED. This proposal is generic and can be extended to create the proposed hybrid entangled state of a DV qudit and a CV qudit, by using two microwave or optical cavities coupled via a natural or artificial three-level atom. This work is a demonstration to produce hybrid entangled states of DV qudits (encoded using Fock states) and CV qudits (encoded with cat states).

There are two additional motivations for this work, which are briefly introduced as follows: (i) Over the past years, a lot of works have been devoted to the hybrid entanglement engineering with DV *qubits* and CV *qubits*. A DV qubit has a two-dimensional Hilbert space formed, e.g., by orthogonal polarizations or the absence and presence of a single photon; while a CV qubit is usually encoded with two quasiorthogonal coherent states or two orthogonal cat states of light. Theoretical schemes have been proposed for producing hybrid entangled states of DV qubits and CV qubits, by means of cavity QED, circuit QED, or linear optical devices [26–32]. Moreover, hybrid entanglement of DV qubits and CV qubits has been experimentally generated in a linear optical system [33,34]. However, after a deep search of literature, we found that the previous works are limited to generation of hybrid entangled states of DV *qubits* and CV *qubits*, while how to prepare hybrid entangled states of DV *qudits* (encoded using Fock states) and CV *qudits* (encoded with cat states) has not been reported yet. (ii) Cat-state qubits, encoded by cat states, have attracted wide attention due to their improved life span [35]. Recently, there has been increasing interest in quantum computing with cat-state qubits. Schemes have been put forward for creating entanglement of multiple cat-state qubits [36,37] and realizing quantum gates for a single cat-state qubit [38,39], two cat-state qubits [27,40], and multiple cat-state qubits [41]. Furthermore, quantum gates of a single cat-state qubit [42] and Bell states of two cat-state qubits [43] have been experimentally demonstrated. However, there is no study on construction of *qudits* using cat states or generation of entangled states of such qudits.

We stress that this work is different from Ref. [32] (where the PO qubits are DV qubits while the WO qubits are CV qubits). First, Ref. [32] is on the generation of hybrid Greenberger-Horne-Zeilinger entangled states of DV and CV qubits, thus the focus of [32] is on qubits (two-state quantum systems); however, this work is on the creation of hybrid entanglement between a DV qudit and a CV qudit, thus it focuses on qudits (high-dimensional quantum systems). Second, the CV qubits in Ref. [32] are encoded with coherent states while the CV qudit in this work is encoded with cat states, hence the encoding is different for the two works. Lastly, Ref. [32] does not use a two-photon process while this work employs a two-photon process, thus the physical mechanisms used in the two works are also different.

This paper is organized as follows. In Sec. II, we show the construction of a CV qudit using cat states and prove the condition necessary for any two of cat states to be quasiorthogonal to each other. In Sec. III, we explicitly show how to generate the hybrid maximally entangled state of a DV qudit and a CV qudit based on circuit QED. In Sec. IV, we numerically analyze the experimental feasibility for creating a hybrid maximally entangled state of a DV qutrit and a CV qutrit. A conclusion is presented in Sec. V.

## II. CONSTRUCTION OF A CV QUDIT USING CAT STATES

A CV qudit is encoded using $d$ quasiorthogonal cat states $\{|C_0\rangle, |C_1\rangle, |C_2\rangle, \ldots, |C_{d-1}\rangle\}$ of a bosonic mode, where the $n$th cat state is expressed as $|C_n\rangle = |\alpha e^{in\phi}\rangle + |-\alpha e^{in\phi}\rangle$ with $n \in \{0, 1, 2, \ldots, d-1\}$. Assume that the photon annihilation operator of the bosonic mode is denoted as $\hat{a}$. One can easily find $\hat{a}^2|C_n\rangle = \alpha^2 e^{i2n\phi}|C_n\rangle$. This equality implies that the cat state $|C_n\rangle$ is an eigenstate of the square of the photon annihilation operator with an eigenvalue $\alpha^2 e^{i2n\phi}$, which can vary continuously because the parameter $\alpha$ or $\phi$ is a continuous variable. Since the eigenvalue here varies continuously, a qudit encoded with cat states is called a CV qudit. This is similar to the situation that a qubit (qudit) encoded with coherent states is often called as a CV qubit (CV qudit) because one has $\hat{a}|\alpha\rangle = \alpha|\alpha\rangle$ for the coherent state $|\alpha\rangle$, which indicates that the coherent state $|\alpha\rangle$ is an eigenstate of the photon annihilation operator with an eigenvalue $\alpha$ that can change continuously.

To make any two of cat states \textrm{quasiorthogonal} to each other, here we adopt a good condition

$$|\langle C_m | C_n \rangle|^2 < 4 \times 10^{-4} \approx 0, \quad (1)$$

for all $n \neq m$ ($m, n \in \{0, 1, 2, \ldots, d-1\}$). In the following, we will give a discussion on the choice of $\alpha$ and $\phi$ required by Eq. (1).

In terms of $|C_n\rangle = |\alpha e^{in\phi}\rangle + |-\alpha e^{in\phi}\rangle$ and $|C_m\rangle = |\alpha e^{im\phi}\rangle + |-\alpha e^{im\phi}\rangle$, the left side of Eq. (1) can be written as

$$|\langle C_m | C_n \rangle|^2 = 4(e^{-A|\alpha|^2} + e^{-B|\alpha|^2} + 2e^{-2|\alpha|^2} \cos[2|\alpha|^2 \sin(n-m)\phi]), \quad (2)$$

where

$$A = |1 - e^{i(n-m)\phi}|^2 = 2 - 2\cos(n-m)\phi,$$
$$B = |1 + e^{i(n-m)\phi}|^2 = 2 + 2\cos(n-m)\phi. \quad (3)$$

Let us first assume that $\alpha$ is chosen to satisfy the following inequality:

$$|\alpha| \geqslant \sqrt{10}/\sin(\pi/d), \quad (4)$$

from which we have $\pi/d \geqslant \arcsin(\sqrt{10}/|\alpha|)$. Define $\theta = \arcsin(\sqrt{10}/|\alpha|)$. Thus, we have $\pi/d \geqslant \theta$, which can be further written as $\theta \leqslant (\pi - \theta)/(d - 1)$. Next, we assume that $\phi$

is chosen to meet the following inequality:

$$\theta \leqslant |\phi| \leqslant (\pi - \theta)/(d-1). \quad (5)$$

Because of the inequality (5) and $1 \leqslant |n - m| \leqslant d - 1$ ($n \neq m$), one has

$$\frac{\theta}{|n-m|} \leqslant \theta \leqslant |\phi| \leqslant \frac{\pi - \theta}{d-1} \leqslant \frac{\pi - \theta}{|n-m|}$$

$$\Rightarrow \theta \leqslant |(n-m)\phi| \leqslant \pi - \theta$$

$$\Rightarrow |\sin(n-m)\phi| \geqslant \sin\theta = \frac{\sqrt{10}}{|\alpha|}$$

$$\Rightarrow \sin^2(n-m)\phi \geqslant \frac{10}{|\alpha|^2}$$

$$\Rightarrow 1 - \sin^2(n-m)\phi \leqslant 1 - \frac{10}{|\alpha|^2}$$

$$< 1 - \frac{10}{|\alpha|^2} + \frac{25}{|\alpha|^4}$$

$$\Rightarrow \cos^2(n-m)\phi < \left(1 - \frac{5}{|\alpha|^2}\right)^2$$

$$\Rightarrow -\left(1 - \frac{5}{|\alpha|^2}\right) < \cos(n-m)\phi$$

$$< \left(1 - \frac{5}{|\alpha|^2}\right). \quad (6)$$

According to the expression of $A$ and $B$ given above and the inequality (6), one has

$$\begin{aligned} 10 < & A|\alpha|^2 < 4|\alpha|^2 - 10, \\ 10 < & B|\alpha|^2 < 4|\alpha|^2 - 10, \end{aligned} \quad (7)$$

which result in

$$\begin{aligned} e^{-A|\alpha|^2} < e^{-10} < 0.454 \times 10^{-4}, \\ e^{-B|\alpha|^2} < e^{-10} < 0.454 \times 10^{-4}. \end{aligned} \quad (8)$$

From the inequality $|\alpha| \geqslant \sqrt{10}/\sin(\pi/d)$, we have $2|\alpha|^2 \geqslant 20/\sin^2 \pi/d \geqslant 20$, resulting in

$$e^{-2|\alpha|^2} \leqslant e^{-20} < 0.21 \times 10^{-8}. \quad (9)$$

Thus, according to Eqs. (2), (8), and (9), we obtain

$$\begin{aligned} |\langle C_m | C_n \rangle|^2 &= 4(e^{-A|\alpha|^2} + e^{-B|\alpha|^2} \\ &\quad + 2e^{-2|\alpha|^2} \cos[2|\alpha|^2 \sin(n-m)\phi]) \\ &\leqslant 4(e^{-A|\alpha|^2} + e^{-B|\alpha|^2} + 2e^{-2|\alpha|^2}) \\ &< 4 \times 10^{-4}. \end{aligned} \quad (10)$$

In the above, we have proved that the condition (1) can be achieved if $\alpha$ and $\phi$ are chosen to satisfy the inequalities (4) and (5). For simplicity, we choose $\alpha = \sqrt{10}/\sin(\pi/sd)$ and $|\phi| = \pi/sd$, with $s \geqslant 1$. In this case, one has $\pi/sd = \arcsin(\sqrt{10}/|\alpha|) = \theta = |\phi|$. It is easy to verify that the choice of $\alpha$ and $\phi$ here satisfies the inequalities (4) and (5). Note that the choice of $\alpha = \sqrt{10}/\sin(\pi/sd)$ and $|\phi| = \pi/sd$ (with $s \geqslant 1$) will be applied in the next sections.

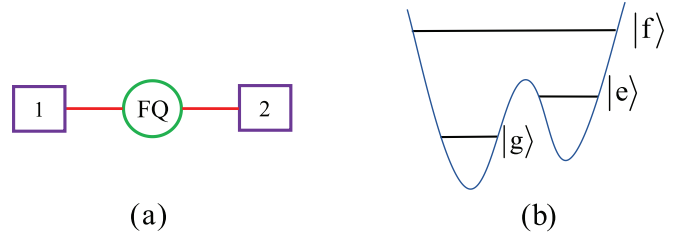

FIG. 1. (a) Schematic diagram of two microwave cavities coupled to a flux qutrit (FQ). Each square represents a one-dimensional or three-dimensional cavity. The circle in the middle represents the flux qutrit, which is capacitively or inductively coupled to each cavity. (b) The three levels of the flux qutrit form a $\Lambda$-type level structure, where the transition between the two lowest levels can be made weak by increasing the barrier between the two potential wells.

## III. GENERATION OF A MAXIMALLY ENTANGLED STATE OF A DV QUDIT AND A CV QUDIT

Let us now consider a setup consisting of two microwave cavities coupled to a superconducting flux qutrit [Fig. 1(a)]. The flux qutrit has a $\Lambda$-type three levels $|g\rangle$, $|e\rangle$, and $|f\rangle$ [Fig. 1(b)]. Suppose that the transition between the two lowest levels $|g\rangle$ and $|e\rangle$ is weak, which can be made by increasing the barrier between the two potential wells [Fig. 1(b)]. Assume that cavity 1 is initially in a superposition of Fock states $\frac{1}{\sqrt{d}}\sum_{n=0}^{d-1}|n\rangle$, where $|n\rangle$ is a Fock state with $n$ photons. In addition, assume that cavity 2 is initially in a cat state $|\alpha\rangle + |-\alpha\rangle$. The coupler qutrit is initially decoupled from the two cavities and is in the ground state $|g\rangle$. Thus, the initial state of the whole system is given by

$$|\psi\rangle_1 = \frac{1}{\sqrt{d}} \sum_{n=0}^{d-1} |n\rangle \otimes (|\alpha\rangle + |-\alpha\rangle)|g\rangle. \quad (11)$$

The state preparation includes two basic operations, which are described below:

Step (i): Adjust the frequency of cavity 1 such that cavity 1 is dispersively coupled to the $|g\rangle \leftrightarrow |f\rangle$ transition with coupling constant $g_1$ and detuning $\Delta_1$ but highly detuned (decoupled) from the $|e\rangle \leftrightarrow |f\rangle$ transition of the qutrit [Fig. 2(a)]. Meanwhile, adjust the frequency of cavity 2 such that cavity 2 is dispersively coupled to the $|e\rangle \leftrightarrow |f\rangle$ transition with coupling constant $g_2$ and detuning $\Delta_2$ but highly detuned (decoupled) from the $|g\rangle \leftrightarrow |f\rangle$ transition of the qutrit [Fig. 2(a)]. The coupling of each cavity with the $|g\rangle \leftrightarrow |e\rangle$ transition of the qutrit is negligible because the $|g\rangle \leftrightarrow |e\rangle$ transition is forbidden or weak. In the interaction picture and after making the rotating-wave approximation (RWA), the Hamiltonian is given by (in units of $\hbar = 1$)

$$H_{I,1} = g_1(e^{i\Delta_1 t} \hat{a}_1^\dagger \sigma_{fg}^- + \text{H.c.}) + g_2(e^{i\Delta_2 t} \hat{a}_2^\dagger \sigma_{fe}^- + \text{H.c.}), \quad (12)$$

where $\hat{a}_1^\dagger$ ($\hat{a}_2^\dagger$) is the photon creation operator of cavity 1 (2), $\Delta_1 = \omega_{c_1} - \omega_{fg} > 0$, $\Delta_2 = \omega_{c_2} - \omega_{fe} > 0$, $\sigma_{fg}^- = |g\rangle\langle f|$, $\sigma_{fe}^- = |e\rangle\langle f|$. Here, $\Delta_1 = \Delta_2 + \delta$ with $\delta = \omega_{c_1} - \omega_{c_2} - \omega_{eg} > 0$ [Fig. 2(a)], $\omega_{c_1}$ ($\omega_{c_2}$) is the frequency of cavity 1 (2), while $\omega_{fe}$, $\omega_{eg}$, and $\omega_{fg}$ are the $|e\rangle \leftrightarrow |f\rangle$, $|g\rangle \leftrightarrow |e\rangle$, and $|g\rangle \leftrightarrow |f\rangle$ transition frequencies of the qutrit, respectively.

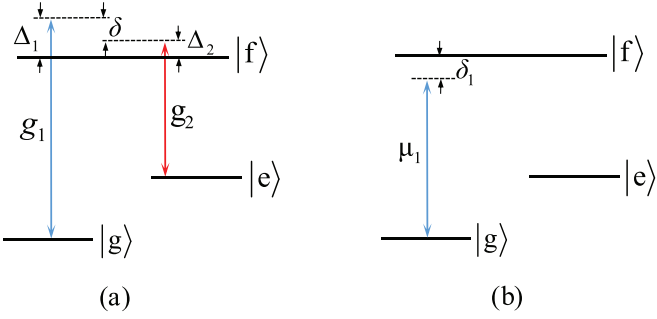

FIG. 2. (a) Cavity 1 is dispersively coupled to the $|g\rangle \leftrightarrow |f\rangle$ transition with coupling constant $g_1$ and detuning $\Delta_1 = \omega_{c_1} - \omega_{fg} > 0$, while cavity 2 is dispersively coupled to the $|e\rangle \leftrightarrow |f\rangle$ transition with coupling constant $g_2$ and detuning $\Delta_2 = \omega_{c_2} - \omega_{ef} > 0$. (b) Cavity 1 is dispersively coupled to the $|g\rangle \leftrightarrow |f\rangle$ transition with coupling constant $\mu_1$ and detuning $\delta_1 = \omega_{fg} - \widetilde{\omega}_{c_1} > 0$, while the frequency of cavity 2 is adjusted such that cavity 2 is decoupled from the qutrit. Here, $\widetilde{\omega}_{c_1}$ is the adjusted frequency of cavity 1.

In the large-detuning regime of $\Delta_1 \gg g_1$ and $\Delta_2 \gg g_2$, the Hamiltonian (12) becomes [44–46]

$$H_{e,1} = \lambda_1(\hat{a}_1^\dagger \hat{a}_1 \sigma_{gg} - \hat{a}_1 \hat{a}_1^\dagger \sigma_{ff}) + \lambda_2(\hat{a}_2^\dagger \hat{a}_2 \sigma_{ee} - \hat{a}_2 \hat{a}_2^\dagger \sigma_{ff})$$
$$+ \lambda(e^{i\delta t} \hat{a}_1^\dagger \hat{a}_2 \sigma_{eg}^- + \text{H.c.}), \quad (13)$$

where the terms in the first two lines describe the photon-number-dependent Stark shifts of the energy levels $|g\rangle$, $|e\rangle$, and $|f\rangle$, while the terms in the last line describe the $|g\rangle \leftrightarrow |e\rangle$ coupling, which is caused due to the cooperation of cavities 1 and 2 through a *two-photon process*. In addition, $\sigma_{gg} = |g\rangle\langle g|$, $\sigma_{ee} = |e\rangle\langle e|$, $\sigma_{ff} = |f\rangle\langle f|$, $\sigma_{eg}^- = |g\rangle\langle e|$, $\lambda_1 = g_1^2/\Delta_1$, $\lambda_2 = g_2^2/\Delta_2$, and $\lambda = (g_1 g_2/2)(1/\Delta_1 + 1/\Delta_2)$. For the derivation of the Hamiltonian (13), please see the Appendix.

If $\delta \gg \{\lambda_1, \lambda_2, \lambda\}$, the Hamiltonian $H_{e,1}$ can be further expressed as [44–46]

$$H_{e,1} = \lambda_1(\hat{a}_1^\dagger \hat{a}_1 \sigma_{gg} - \hat{a}_1 \hat{a}_1^\dagger \sigma_{ff}) + \lambda_2(\hat{a}_2^\dagger \hat{a}_2 \sigma_{ee} - \hat{a}_2 \hat{a}_2^\dagger \sigma_{ff})$$
$$+ \chi(\hat{a}_1^\dagger \hat{a}_1 \hat{a}_2 \hat{a}_2^\dagger \sigma_{gg} - \hat{a}_1 \hat{a}_1^\dagger \hat{a}_2^\dagger \hat{a}_2 \sigma_{ee}), \quad (14)$$

where $\chi = \lambda^2/\delta$. The derivation of the Hamiltonian (14) can be found in the Appendix.

Note that the Hamiltonian (14) does not induce both $|g\rangle \to |e\rangle$ and $|g\rangle \to |f\rangle$ transitions. Thus, as long as the levels $|e\rangle$ and $|f\rangle$ are initially unpopulated, they will remain not occupied. In this case, the Hamiltonian (14) reduces to $H_{e,1} = \lambda_1 \hat{a}_1^\dagger \hat{a}_1 \sigma_{gg} + \chi \hat{a}_1^\dagger \hat{a}_1 \hat{a}_2 \hat{a}_2^\dagger \sigma_{gg}$, which can be rewritten as

$$H_{e,1} = (\lambda_1 + \chi)\hat{a}_1^\dagger \hat{a}_1 \sigma_{gg} + \chi(\hat{a}_1^\dagger \hat{a}_1)(\hat{a}_2^\dagger \hat{a}_2)\sigma_{gg}, \quad (15)$$

where we have used $\hat{a}_2 \hat{a}_2^\dagger = 1 + \hat{a}_2^\dagger \hat{a}_2$ because of $[\hat{a}_2, \hat{a}_2^\dagger] = 1$. Under this Hamiltonian, for an interaction time $\tau$, the state (11) changes to

$$|\psi\rangle_2 = \frac{1}{\sqrt{d}} \sum_{n=0}^{d-1} e^{-i(\lambda_1 + \chi)n\tau} |n\rangle$$
$$\otimes (|\alpha e^{-i\chi n\tau}\rangle + |-\alpha e^{-i\chi n\tau}\rangle)|g\rangle. \quad (16)$$

Step (ii): Adjust the frequency of cavity 1 such that cavity 1 is dispersively coupled to the $|g\rangle \leftrightarrow |f\rangle$ transition of the qutrit with coupling constant $\mu_1$ and detuning $\delta_1$ [Fig. 2(b)]. In addition, adjust the frequency of cavity 2 such that cavity 2 is decoupled from the qutrit. Under these considerations, the Hamiltonian in the interaction picture and after making the RWA is given by

$$H_{I,2} = \mu_1(e^{-i\delta_1 t} \hat{a}_1^\dagger \sigma_{fg}^- + \text{H.c.}), \quad (17)$$

where $\delta_1 = \omega_{fg} - \widetilde{\omega}_{c_1} > 0$ and $\widetilde{\omega}_{c_1}$ is the adjusted frequency of cavity 1. For $\delta_1 \gg \mu_1$, the Hamiltonian $H_{I,2}$ becomes [45,46]

$$H_{e,2} = -\widetilde{\lambda}_1 \hat{a}_1^\dagger \hat{a}_1 \sigma_{gg} + \widetilde{\lambda}_1 \hat{a}_1 \hat{a}_1^\dagger \sigma_{ff}, \quad (18)$$

where $\widetilde{\lambda}_1 = \mu_1^2/\delta_1$. When the level $|f\rangle$ is not occupied, the Hamiltonian (18) is reduced to

$$H_{e,2} = -\widetilde{\lambda}_1 \hat{a}_1^\dagger \hat{a}_1 \sigma_{gg}. \quad (19)$$

Under this Hamiltonian and for an interaction time $\tau$, the state (16) becomes

$$|\psi\rangle_3 = \left[ \frac{1}{\sqrt{d}} \sum_{n=0}^{d} e^{-i(\lambda_1 + \chi - \widetilde{\lambda}_1)n\tau} |n\rangle (|\alpha e^{-i\chi n\tau}\rangle \right.$$
$$\left. + |-\alpha e^{-i\chi n\tau}\rangle) \right] |g\rangle. \quad (20)$$

In the following, we set $\lambda_1 + \chi = \widetilde{\lambda}_1$. Thus, the state (20) becomes

$$|\psi\rangle_4 = \left[ \frac{1}{\sqrt{d}} \sum_{n=0}^{d-1} |n\rangle \otimes (|\alpha e^{in\phi}\rangle + |-\alpha e^{in\phi}\rangle) \right] |g\rangle, \quad (21)$$

where we have defined $\phi = -\chi \tau$. It is obvious that the two-cavity state involved in Eq. (21) can be expressed as

$$|\psi\rangle_5 = \frac{1}{\sqrt{d}} (|0\rangle|C_0\rangle + |1\rangle|C_1\rangle + |2\rangle|C_2\rangle$$
$$+ \cdots + |d-1\rangle|C_{(d-1)}\rangle), \quad (22)$$

where $|C_n\rangle = |\alpha e^{in\phi}\rangle + |-\alpha e^{in\phi}\rangle$ is a cat state ($n = 0, 1, 2, \ldots, d-1$).

For two qudits, the maximally entangled state takes the form of $\frac{1}{\sqrt{d}} \sum_{j=0}^{d-1} |j\rangle_1 |j\rangle_2$ [47,48], where the subscripts 1 and 2 represent the two qudits, the state $|j\rangle_1$ ($|j\rangle_2$) is the basis state of the first (second) qudit, and any two of the $d$ basis states of each qudit are orthogonal to each other. Let us now consider that a DV qudit is encoded with cavity 1's Fock states $\{|0\rangle, |1\rangle, |2\rangle, \ldots, |d-1\rangle\}$, while a CV qudit is encoded with cavity 2's cat states $\{|C_0\rangle, |C_1\rangle, |C_2\rangle, \ldots, |C_n\rangle\}$. From the discussion here, one can see that, to ensure that the state (22) is a hybrid maximally entangled state of a DV qudit and a CV qudit, any two of the $d$ cat states $\{|C_0\rangle, |C_1\rangle, |C_2\rangle, \ldots, |C_n\rangle\}$ of the CV qudit require to be orthogonal. In the previous section, we have proved that for a sufficiently large $\alpha$ and by choosing an appropriate $\phi$, any two of the cat states can be made to be quasiorthogonal to each other. Therefore, as long as any two of the cat states are quasiorthogonal to each other, the state (22) is a hybrid maximally entangled state of a DV qudit and a CV qudit to a very good approximation.

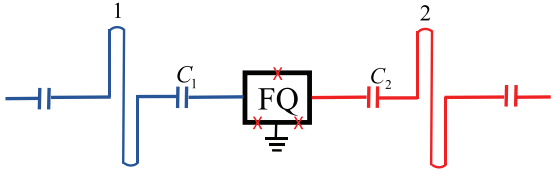

FIG. 3. Diagram of two 1D microwave cavities capacitively coupled to a superconducting flux qutrit (FQ). Each cavity here is a one-dimensional transmission line resonator. The flux qutrit consists of three Josephson junctions and a superconducting loop.

From the above description, one can see the following:

(i) The coupler qutrit remains in the ground state during the entire operation. Therefore, decoherence from the qutrit is greatly suppressed.

(ii) Neither measurement on the state of the coupler qutrit nor measurement on the state of the cavities is needed. Thus, the preparation of the entangled state (22) is deterministic.

(iii) The above condition $\lambda_1 + \chi = \tilde{\lambda}_1$ turns out into $\delta_1 = \mu_1^2 [g_1^2/\Delta_1 + \frac{g_1 g_2}{2}(\frac{1}{\Delta_1} + \frac{1}{\Delta_2})]^{-1}$, which results in

$$\tilde{\omega}_{c_1} = \frac{\omega_{fg}}{1 - \frac{g_1^2}{\omega_{c_1}} \left[ g_1^2/\Delta_1 + \frac{g_1 g_2}{2}(\frac{1}{\Delta_1} + \frac{1}{\Delta_2}) \right]^{-1}}. \quad (23)$$

Equation (23) can be met by carefully selecting the cavity frequency $\tilde{\omega}_{c_1}$, given $\omega_{c_1}$, $g_1$, $g_2$, $\Delta_1$, and $\Delta_2$. In deriving Eq. (23), we have used $\mu_1/g_1 = \sqrt{\tilde{\omega}_{c_1}/\omega_{c_1}}$ because of $g_1 = c\sqrt{\omega_{c_1}}\phi_{gf}$ and $\mu_1 = c\sqrt{\tilde{\omega}_{c_1}}\phi_{gf}$ [49], where $\phi_{gf}$ represents the dipole coupling matrix element between the two levels $|g\rangle$ and $|f\rangle$ of the flux qutrit, and $c$ is a constant.

(iv) The frequency of a microwave cavity can be quickly tuned in $1 \sim 2$ ns [50,51]. In fact, adjusting the cavity frequency is unnecessary. Alternatively, the coupling or decoupling of the qutrit with the cavities can be achieved by adjusting the level spacings of the qutrit. For a superconducting device, the level spacings can be rapidly adjusted within 1–3 ns [52,53].

(v) In the above, we have set $\phi = -\chi \tau$. Thus, by adopting $|\phi| = \pi/sd$ ($s \geqslant 1$) as introduced in Sec. II, we have

$$\tau = \frac{\pi}{sd\chi}. \quad (24)$$

Note that $\tau$ is the operation time required for each of steps (i) and (ii), as described in the previous section. Hence, the total operation time required for preparing the hybrid entangled state is $2\tau = \frac{2\pi}{sd\chi}$, which *decreases* as the dimension size $d$ of each qudit increases.

## IV. EXPERIMENTAL FEASIBILITY

As an example of our proposal, we now give an investigation on the experimental feasibility for generating the hybrid entangled state of a DV qutrit and a CV qutrit (i.e., the case for $d = 3$), by considering a setup of a superconducting flux qutrit coupled to two one-dimensional (1D) microwave cavities or resonators (Fig. 3). In the following, we first give a brief review on the experimental preparation of the initial state given in Eq. (11), we will introduce the full Hamiltonians used in our numerical simulations, and then present our numerical results. Finally, we will give a brief discussion.

### A. Preparation of the initial state

Let us first give a brief discussion on the possibility of experimentally creating a superposition of Fock states $\frac{1}{\sqrt{d}} \sum_{n=0}^{d-1} |n\rangle$ and a cat state $|\alpha\rangle + |-\alpha\rangle$, which are two ingredients of the initial state for the generation of the hybrid entangled state (22). Over the past years, much progress has been achieved on the experimental creations of a superposition of Fock states and cat states based on circuit QED. For instances, the experimental creation of a superposition of Fock states with the first 3 Fock states (i.e., $|0\rangle$, $|1\rangle$, $|2\rangle$) [54] or the first 10 Fock states [55] has been reported. In addition, Ref. [56] reported the experimental preparation of a cat state $|\alpha\rangle \pm |-\alpha\rangle$ with an average photon number $\bar{n} = 2.6$ (i.e., an amplitude $|\alpha| \approx 1.61$); Ref. [57] reported the experimental preparation of an atom-cavity entangled Bell state $(|g\rangle|\beta\rangle + |e\rangle|-\beta\rangle)/\sqrt{2}$ with $|\beta| \leqslant 2$, based on which a cat state $|\beta\rangle \pm |-\beta\rangle$ with $|\beta| \leqslant 2$ can be directly created by a measurement on the atom along a rotated basis $|\pm\rangle = (|g\rangle \pm |e\rangle)/\sqrt{2}$. Furthermore, Ref. [58] reported the experimental creation of a cat state $|\beta\rangle + |-\beta\rangle$ with the size of $S = 111$ photons. According to $S = 4|\beta|^2 = 111$ (see the Supplemental Material of [58]), the cat state reported in [58] has an amplitude $|\beta| \leqslant 5.27$. The discussion here implies that a superposition of Fock states with $d \leqslant 10$ and a cat state with $|\alpha| \leqslant 5.27$ can be prepared within current circuit QED experiments. In addition, with development of circuit QED technology, the experimental preparation of a superposition of Fock states with a larger $d$ and a cat state with a larger amplitude $|\alpha|$ is also possible in the near future.

### B. Full Hamiltonians

From the description in the previous section, it can be seen that the entangled state preparation involves the following two basic operations:

(i) The first operation is characterized by the Hamiltonian (12). When the intercavity crosstalk between the two cavities and the unwanted coupling of each cavity with the irrelevant level transitions of the qutrit are considered, the Hamiltonian $H_{I,1}$ given in Eq. (12) is modified as $\tilde{H}_{I,1} = H_{I,1} + \varepsilon_1$, with

$$\begin{aligned} \varepsilon_1 &= \tilde{g}_1(e^{i\tilde{\Delta}_1 t}\hat{a}_1^\dagger \sigma_{fe}^- + \text{H.c.}) + g_1'(e^{i\Delta_1' t}\hat{a}_1^\dagger \sigma_{eg}^- + \text{H.c.}) \\ &+ \tilde{g}_2(e^{-i\tilde{\Delta}_2 t}\hat{a}_2^\dagger \sigma_{fg}^- + \text{H.c.}) + g_2'(e^{i\Delta_2' t}\hat{a}_2^\dagger \sigma_{eg}^- + \text{H.c.}) \\ &+ g_{12}(e^{i\Delta_{12} t}\hat{a}_1^\dagger \hat{a}_2 + \text{H.c.}), \end{aligned} \quad (25)$$

where the first term in line one represents the unwanted coupling between cavity 1 and the $|e\rangle \leftrightarrow |f\rangle$ transition of the qutrit with coupling constant $\tilde{g}_1$ and detuning $\tilde{\Delta}_1 = \omega_{c_1} - \omega_{fe} > 0$, while the second term in line one represents the unwanted coupling between cavity 1 and the $|g\rangle \leftrightarrow |e\rangle$ transition with coupling constant $g_1'$ and detuning $\Delta_1' = \omega_{c_1} - \omega_{eg} > 0$ [Fig. 4(a)]; the first term in line two represents the unwanted coupling between cavity 2 and the $|g\rangle \leftrightarrow |f\rangle$ transition with coupling constant $\tilde{g}_2$ and detuning $\tilde{\Delta}_2 = \omega_{fg} - \omega_{c_2} > 0$, while the second term in line two represents the unwanted coupling between cavity 2 and the $|g\rangle \leftrightarrow |e\rangle$ transition with coupling constant $g_2'$ and detuning $\Delta_2' = \omega_{c_2} - \omega_{eg} > 0$ [Fig. 4(a)]; the last line describes the intercavity crosstalk

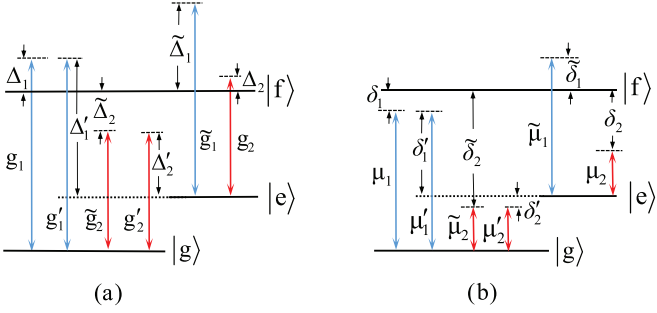

FIG. 4. (a) Illustration of the unwanted coupling between cavity 1 and the $|e\rangle \leftrightarrow |f\rangle$ transition with coupling constant $\widetilde{g}_1$ and detuning $\widetilde{\Delta}_1 = \omega_{c_1} - \omega_{fe} > 0$ as well as the unwanted coupling between cavity 1 and the $|g\rangle \leftrightarrow |e\rangle$ transition with coupling constant $g'_1$ and detuning $\Delta'_1 = \omega_{c_1} - \omega_{eg} > 0$. Illustration of the unwanted coupling between cavity 2 and the $|g\rangle \leftrightarrow |f\rangle$ transition with coupling constant $\widetilde{g}_2$ and detuning $\widetilde{\Delta}_2 = \omega_{fg} - \omega_{c_2} > 0$ as well as the unwanted coupling between cavity 2 and the $|g\rangle \leftrightarrow |e\rangle$ transition with coupling constant $g'_2$ and detuning $\Delta'_2 = \omega_{c_2} - \omega_{eg} > 0$. (b) Illustration of the unwanted coupling between cavity 1 and the $|e\rangle \leftrightarrow |f\rangle$ transition with coupling constant $\widetilde{\mu}_1$ and detuning $\widetilde{\delta}_1 = \widetilde{\omega}_{c_1} - \omega_{fe} > 0$ as well as the unwanted coupling between cavity 1 and the $|g\rangle \leftrightarrow |e\rangle$ transition with coupling constant $\mu'_1$ and detuning $\delta'_1 = \widetilde{\omega}_{c_1} - \omega_{eg} > 0$. Illustration of the unwanted coupling between cavity 2 and the $|e\rangle \leftrightarrow |f\rangle$ transition with coupling constant $\mu_2$ and detuning $\delta_2 = \omega_{fe} - \widetilde{\omega}_{c_2} > 0$, the unwanted coupling between cavity 2 and the $|g\rangle \leftrightarrow |f\rangle$ transition with coupling constant $\widetilde{\mu}_2$ and $\widetilde{\delta}_2 = \omega_{fg} - \widetilde{\omega}_{c_2} > 0$, as well as the unwanted coupling between cavity 2 and the $|g\rangle \leftrightarrow |e\rangle$ transition with coupling constant $\mu'_2$ and detuning $\delta'_2 = \omega_{eg} - \widetilde{\omega}_{c_2} > 0$.

with coupling strength $g_{12}$ and frequency difference $\Delta_{12} = \omega_{c_1} - \omega_{c_2}$.

(ii) The second operation is described by the Hamiltonian (17). When the intercavity crosstalk between the two cavities and the unwanted coupling of each cavity with the irrelevant level transitions are taken into account, the Hamiltonian $H_{I,2}$ given in Eq. (17) is modified as $\widetilde{H}_{I,2} = H_{I,2} + \varepsilon_2$, with

$$\varepsilon_2 = \widetilde{\mu}_1(e^{i\widetilde{\delta}_1 t}\hat{a}_1^\dagger \sigma_{fe}^- + \text{H.c.}) + \mu'_1(e^{i\delta'_1 t}\hat{a}_1^\dagger \sigma_{eg}^- + \text{H.c.})$$
$$+ \mu_2(e^{-i\delta_2 t}\hat{a}_2^\dagger \sigma_{fe}^- + \text{H.c.}) + \widetilde{\mu}_2(e^{-i\widetilde{\delta}_2 t}\hat{a}_2^\dagger \sigma_{fg}^- + \text{H.c.})$$
$$+ \mu'_2(e^{-i\delta'_2 t}\hat{a}_2^\dagger \sigma_{eg}^- + \text{H.c.}) + \widetilde{g}_{12}(e^{i\widetilde{\Delta}_{12}t}\hat{a}_1^\dagger \hat{a}_2 + \text{H.c.}),$$
(26)

where the first term in line one represents the unwanted coupling between cavity 1 and the $|e\rangle \leftrightarrow |f\rangle$ transition with coupling constant $\widetilde{\mu}_1$ and detuning $\widetilde{\delta}_1 = \widetilde{\omega}_{c_1} - \omega_{fe} > 0$, while the second term in line one represents the unwanted coupling between cavity 1 and the $|g\rangle \leftrightarrow |e\rangle$ transition with coupling constant $\mu'_1$ and detuning $\delta'_1 = \widetilde{\omega}_{c_1} - \omega_{eg} > 0$; the first term in line two represents the unwanted coupling between cavity 2 and the $|e\rangle \leftrightarrow |f\rangle$ transition with coupling constant $\mu_2$ and detuning $\delta_2 = \omega_{fe} - \widetilde{\omega}_{c_2} > 0$, the second term in line two represents the unwanted coupling between cavity 2 and the $|g\rangle \leftrightarrow |f\rangle$ transition with coupling constant $\widetilde{\mu}_2$ and detuning $\widetilde{\delta}_2 = \omega_{fg} - \widetilde{\omega}_{c_2} > 0$, while the third term in line two represents the unwanted coupling between cavity 2 and the $|g\rangle \leftrightarrow |e\rangle$ transition with coupling constant $\mu'_2$ and detuning $\delta'_2 = \omega_{eg} - \widetilde{\omega}_{c_2} > 0$ [Fig. 4(b)]; the last line describes the intercavity crosstalk with coupling strength $\widetilde{g}_{12}$ and frequency difference $\widetilde{\Delta}_{12} = \widetilde{\omega}_{c_1} - \widetilde{\omega}_{c_2}$. Here, $\widetilde{\omega}_{c_2}$ is the adjusted frequency of cavity 2.

TABLE I. Parameters used in the numerical simulation. For the definitions of the parameters, please refer to the text. Here, the frequency $\widetilde{\omega}_{c_1}$ is calculated according to Eq. (23). In addition, the coupling constants $\mu_1$ and $\mu_2$ are calculated based on $\mu_1 = \sqrt{\widetilde{\omega}_{c_1}/\omega_{c_1}}g_1$ and $\mu_2 = \sqrt{\widetilde{\omega}_{c_2}/\omega_{c_2}}g_2$.

| | | |
|---|---|---|
| $\omega_{fg}/2\pi = 12.0\,\text{GHz}$ | $\omega_{fe}/2\pi = 7.0\,\text{GHz}$ | $\omega_{eg}/2\pi = 5.0\,\text{GHz}$ |
| $\omega_{c_1}/2\pi = 15.0\,\text{GHz}$ | $\omega_{c_2}/2\pi = 9.0\,\text{GHz}$ | $\widetilde{\omega}_{c_1}/2\pi = 9.96\,\text{GHz}$ |
| $\widetilde{\omega}_{c_2}/2\pi = 4.0\,\text{GHz}$ | $\Delta_1/2\pi = 3.0\,\text{GHz}$ | $\Delta_2/2\pi = 2.0\,\text{GHz}$ |
| $\widetilde{\Delta}_1/2\pi = 8.0\,\text{GHz}$ | $\widetilde{\Delta}_2/2\pi = 3.0\,\text{GHz}$ | $\Delta'_1/2\pi = 10.0\,\text{GHz}$ |
| $\Delta'_2/2\pi = 4.0\,\text{GHz}$ | $\Delta_{12}/2\pi = 6.0\,\text{GHz}$ | $\widetilde{\Delta}_{12}/2\pi = 5.96\,\text{GHz}$ |
| $\delta/2\pi = 1.0\,\text{GHz}$ | $\delta_1/2\pi = 2.04\,\text{GHz}$ | $\widetilde{\delta}_1/2\pi = 2.96\,\text{GHz}$ |
| $\delta'_1/2\pi = 4.96\,\text{GHz}$ | $\delta_2/2\pi = 3.0\,\text{GHz}$ | $\widetilde{\delta}_2/2\pi = 8.0\,\text{GHz}$ |
| $\delta'_2/2\pi = 1.0\,\text{GHz}$ | $g_1/2\pi = 236\,\text{MHz}$ | $g_2/2\pi = 223\,\text{MHz}$ |
| $\mu_1/2\pi = 192.1\,\text{MHz}$ | $\mu_2/2\pi = 148.7\,\text{MHz}$ | |

### C. Numerical results

After taking into account the qutrit decoherence and the cavity decay, the dynamics of the lossy system is determined by the following master equation:

$$\frac{d\rho}{dt} = -i[\widetilde{H}_{I,i}, \rho] + \sum_{l=1}^{2}\kappa_l \mathcal{L}[\hat{a}_l]$$
$$+ \gamma_{eg}\mathcal{L}[\sigma_{eg}^-] + \gamma_{fe}\mathcal{L}[\sigma_{fe}^-] + \gamma_{fg}\mathcal{L}[\sigma_{fg}^-]$$
$$+ \gamma_{\varphi,e}(\sigma_{ee}\rho\sigma_{ee} - \sigma_{ee}\rho/2 - \rho\sigma_{ee}/2)$$
$$+ \gamma_{\varphi,f}(\sigma_{ff}\rho\sigma_{ff} - \sigma_{ff}\rho/2 - \rho\sigma_{ff}/2), \quad (27)$$

where $\widetilde{H}_{I,i}$ (with $i = 1, 2$) are the above modified Hamiltonians $\widetilde{H}_{I,1}$ and $\widetilde{H}_{I,2}$, $\kappa_l$ is the decay rate of cavity $l$ ($l = 1, 2$), $\gamma_{eg}$ is the energy relaxation rate for the level $|e\rangle$, $\gamma_{fe}$ ($\gamma_{fg}$) is the energy relaxation rate of the level $|f\rangle$ corresponding to the decay path $|f\rangle \to |e\rangle$ ($|g\rangle$), $\gamma_{\varphi,e}$ ($\gamma_{\varphi,f}$) is the dephasing rate of the level $|e\rangle$ ($|f\rangle$) of the qutrit, and $\mathcal{L}[\xi] = \xi\rho\xi^\dagger - \xi^\dagger\xi\rho/2 - \rho\xi^\dagger\xi/2$ with $\xi = \hat{a}_l, \sigma_{eg}^-, \sigma_{fe}^-, \sigma_{fg}^-$.

The efficiency of the entire operation is evaluated by the fidelity $\mathcal{F} = \sqrt{\langle\psi_{\text{id}}|\rho|\psi_{\text{id}}\rangle}$, where $|\psi_{\text{id}}\rangle$ is the ideal output state given in Eq. (21) (with $d = 3$ for the present example), which is obtained without considering the system dissipation, the intercavity crosstalk and the unwanted couplings), while $\rho$ is the final density operator of the system in the case when the operations are performed in a realistic situation.

For a flux qutrit, the typical transition frequency between neighboring levels can be made as 1 to 20 GHz. As a concrete example, consider the parameters listed in Table I, which are used in our numerical simulations. With appropriate design of the flux qutrit system [59], one can have $\phi_{fg} \sim \phi_{fe} \sim 10\phi_{eg}$, where $\phi_{ij}$ represents the dipole coupling matrix element between the two levels $|i\rangle$ and $|j\rangle$ with $ij \in \{eg, fe, fg\}$. Thus, we have $\widetilde{g}_1 \sim g_1$, $g'_1 \sim 0.1g_1$; $\widetilde{g}_2 \sim g_2$, $g'_2 \sim 0.1g_2$; $\widetilde{\mu}_1 \sim \mu_1$, $\mu'_1 \sim 0.1\mu_1$; and $\widetilde{\mu}_2 \sim \mu_2$, $\mu'_2 \sim 0.1\mu_2$. For the coupling constants listed in Table I, the maximal value $g_{max} = \max\{g_1, g_2, \widetilde{g}_1, \widetilde{g}_2, \mu_1, \mu_2, \widetilde{\mu}_1, \widetilde{\mu}_2\}$ is $2\pi \times 236$ MHz, which is readily achievable in experiments because a coupling strength $\sim 2\pi \times 636$ MHz was reported for a flux qutrit coupled to a 1D microwave cavity [60].

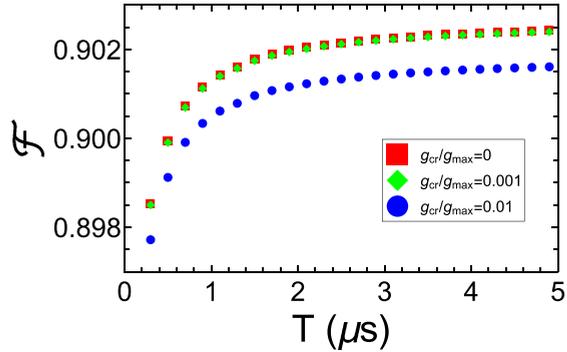

FIG. 5. Fidelity versus $T$ for $g_{cr}/g_{max} = 0$, 0.001, and 0.01. In the numerical simulation, we set $\kappa^{-1} = 10$ $\mu$s. Here, $\kappa$ is the cavity decay rate, and $g_{cr}$ is the crosstalk strength between the two cavities.

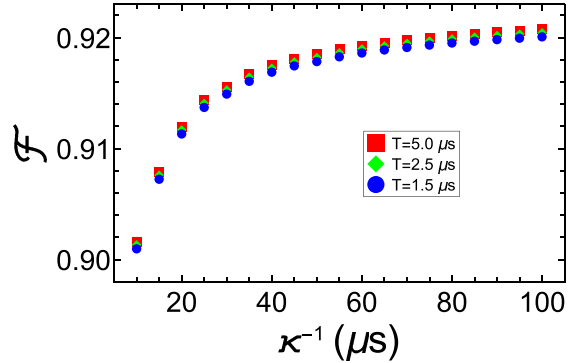

FIG. 6. Fidelity versus $\kappa^{-1}$ for $T = 1.5$, 2.5, and 5.0 $\mu$s. In the numerical simulation, we set $g_{cr} = 0.01 g_{max}$.

In the numerical simulation, other parameters used are (i) $\gamma_{eg}^{-1} = 10T$ $\mu$s, $\gamma_{fe}^{-1} = T$ $\mu$s, $\gamma_{fg}^{-1} = T$ $\mu$s, $\gamma_{\phi e}^{-1} = \gamma_{\phi f}^{-1} = T/2$ $\mu$s, (ii) $\kappa_1 = \kappa_2 = \kappa$, (iii) $s = 1$, $\phi = -\pi/3$, $\alpha = 2\sqrt{10}/3$, (iv) $g_{12} = \widetilde{g}_{12} = g_{cr}$. Because of $\phi_{fg} \sim \phi_{fe} \sim 10\phi_{eg}$, $\gamma_{eg}^{-1}$ is much larger than $\gamma_{fe}^{-1}$ and $\gamma_{fg}^{-1}$. The maximal value of $T$ that we use in the numerical simulations is 5 $\mu$s. For $T = 5$ $\mu$s, the decoherence times $\{\gamma_{eg}^{-1}, \gamma_{fe}^{-1}, \gamma_{fg}^{-1}, \gamma_{\phi e}^{-1}, \gamma_{\phi f}^{-1}\}$ of the qutrit used in the numerical simulations are 2.5–50 $\mu$s. Experimentally, decoherence time 70 $\mu$s to 1 ms for a superconducting flux device has been demonstrated [61,62]. Thus, the decoherence times of the flux qutrit considered in our numerical simulation are a rather conservative case.

To estimate the effect of decoherence from the qutrit, the fidelity versus $T$ for $\kappa^{-1} = 10$ $\mu$s is plotted in Fig. 5 by numerically solving the master equation (27). Figure 5 indicates that when $T \geqslant 1.5$ $\mu$s, the fidelity does not increase much as $T$ increases. This is because the qutrit mostly remains in the ground state during the entire operation and thus decoherence from the higher-energy levels of the qutrit is greatly suppressed. Meanwhile, Fig. 5 shows that the fidelity is slightly affected by the cavity crosstalk when $g_{cr} \leqslant 0.01 g_{max}$ since the fidelity decreases only by 0.001 or less when compared to the fidelity without considering the cavity crosstalk (i.e., the case of $g_{cr} = 0$). This is due to the fact that the frequency detuning between the two cavities is extremely large. For other investigations as depicted in Figs. 6–8, we thus set $g_{cr} = 0.01 g_{max}$, which is achievable in experiments by a prior design of the sample with appropriate capacitances $C_1$ and $C_2$ illustrated in Fig. 3 [63].

Figure 6 illustrates the fidelity versus $\kappa^{-1}$ for $T = 1.5$, 2.5, and 5.0 $\mu$s, which is plotted by numerically solving the master equation (27). From Fig. 6, one can see that the fidelity exceeds 90.1% for $\kappa^{-1} \geqslant 10$ $\mu$s and $T \geqslant 1.5$ $\mu$s. As shown in Fig. 6, the fidelity is sensitive to the cavity decay. This is expected because this work focuses on the preparation of hybrid entangled states of the cavities and thus each cavity is occupied by photons during the entire operation. Through numerical tests, we find that the effect of the cavity decay mainly comes from the cavity (i.e., cavity 2) which hosts cat states.

In reality, the initial state of Eq. (11) may not be prepared perfectly. Thus, we consider a nonideal superposition of Fock states $\mathcal{N}^{-1}[(1/\sqrt{3} + x)|0\rangle + 1/\sqrt{3}|1\rangle + (1/\sqrt{3} - x)|2\rangle]$ with a normalized coefficient $\mathcal{N} = \sqrt{1 + 2x^2}$. For this case, we numerically calculate the fidelity and plot Fig. 7, which shows that the fidelity decreases as the $|x|$ increases. However, for $x \in [-0.06, 0.06]$, i.e., corresponding to a $|x|/(1/\sqrt{3}) \sim 10.4\%$ error in the weights of Fock states, a fidelity greater than 89.5%, 91%, 91.3% can be reached for $\kappa^{-1} = 10, 30, 50$ $\mu$s, respectively.

As shown above, our proposal consists of a two-step operation. To study the effect of the operational time error on the fidelity, we consider the operational time $\tau + \delta\tau$ for the first step while the operational time $\tau - \delta\tau$ for the second step. Namely, the operational time for each step has a small difference $\delta\tau$ compared to the ideal operational time $\tau$ for each step of operations. We numerically calculate the fidelity and plot Fig. 8, which demonstrates that the fidelity decreases by $\sim 0.02$ for $\delta\tau/\tau \in [-0.05, 0.05]$, i.e., corresponding to a 5% error in the typical operation time for each step. This result indicates that the fidelity is sensitive to the operational time error.

The entire operational time is estimated to be $\sim 0.69$ $\mu$s, which is much shorter than the cavity decay time

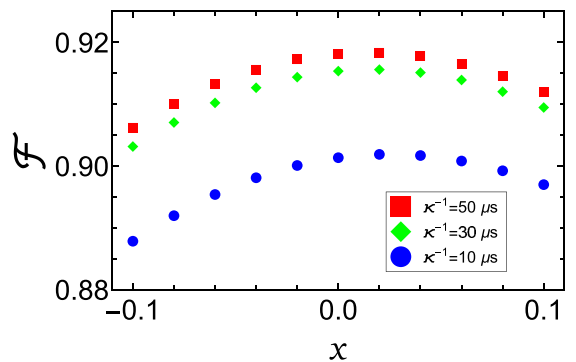

FIG. 7. Fidelity versus $x$ for $\kappa^{-1} = 10, 30,$ and 50 $\mu$s. In the numerical simulation, we set $T = 2.5$ $\mu$s and $g_{cr} = 0.01 g_{max}$.

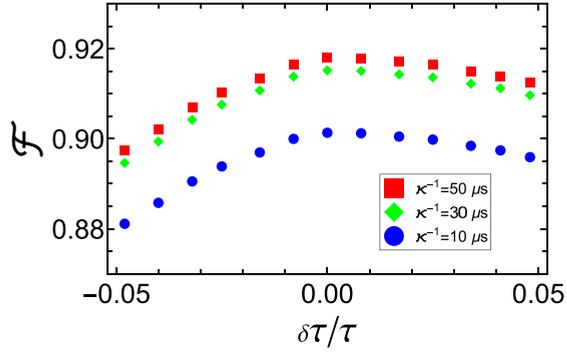

FIG. 8. Fidelity versus $\delta\tau/\tau$ for $\kappa^{-1} = 10$, 30, and 50 $\mu$s. In the numerical simulation, we set $T = 2.5$ $\mu$s and $g_{cr} = 0.01 g_{\max}$.

10–100 $\mu$s used in the numerical simulations. For the cavity frequencies given above and $\kappa^{-1} = 10$ $\mu$s, the cavity quality factors are (i) $Q_1 \sim 9.4 \times 10^5$ for $\omega_{c_1}/2\pi = 15.0$ GHz, $Q_2 \sim 5.65 \times 10^5$ for $\omega_{c_2}/2\pi = 9.0$ GHz; and (ii) $\widetilde{Q}_1 \sim 6.25 \times 10^5$ for $\widetilde{\omega}_{c_1}/2\pi = 9.96$ GHz, $\widetilde{Q}_2 \sim 2.51 \times 10^5$ for $\widetilde{\omega}_{c_2}/2\pi = 4.0$ GHz. The quality factors of the cavities here are available because a high-quality factor $> 10^6$ of a 1D superconducting microwave cavity or resonator was previously reported in experiments [64–66].

### D. Discussion

Our analysis given above demonstrates that the operational fidelity is sensitive to the error in the initial state preparation, the operational time error for each step, and the cavity decay, but it is insensitive to decoherence of the qutrit and the effect of the cavity crosstalk. Thus, to obtain a high fidelity, one would need to have a precise preparation of the initial state, take a good control of the operational time, choose cavities with a high-quality factor, and improve the system parameters. In addition, the fidelity can be increased by choosing a coupler qutrit with a larger level anharmonicity, such that the unwanted couplings of each cavity with the transitions between the irrelevant levels of the qutrit is negligible. We remark that further investigation is needed for each particular experimental setup. However, this requires a rather lengthy and complex analysis, which is beyond the scope of this theoretical work.

## V. CONCLUSION

A continuous-variable (CV) qudit has been constructed using \textrm{quasi}orthogonal cat states of a bosonic mode, when the phase encoded in each cat state is chosen appropriately. Based on the constructed CV qudit and the DV qudit encoded with Fock states, an approach has been proposed to generate the hybrid maximally entangled state of a CV qudit and a DV qudit by using a circuit-QED setup. This proposal relies on the initial preparation of a superposition of Fock states of one cavity and the initial preparation of a cat state of the other cavity. Our proposal for the entanglement generation has the following advantages: (i) After the initial state of each cavity is prepared, the entangled state preparation requires only two basic operations. (ii) The entangled state preparation is deterministic because there is no need of measurement. (iii) The operational time *decreases* as the dimensional size of each qudit increases. (iv) The qutrit remains in the ground state during the entire operation, thus decoherence from the qutrit is greatly suppressed. As an example of this proposal, we have given a discussion on the experimental feasibility for generating the hybrid maximally entangled state of a DV qutrit and a CV qutrit based on circuit QED. This proposal is generic and can be extended to accomplish the same task, by using two microwave or optical cavities coupled to a natural or artificial three-level atom.

## ACKNOWLEDGMENTS

This work was partly supported by the Key-Area Research and Development Program of GuangDong province (Grant No. 2018B030326001), the National Natural Science Foundation of China (NSFC) (Grants No. 11074062, No. 11374083, No. 11504075, No. 11774076, No. 12004253), and the Jiangxi Natural Science Foundation (Grant No. 20192ACBL20051).

## APPENDIX

In this Appendix, we will derive an effective Hamiltonian, based on which we then give a derivation of the Hamiltonian (13) and the Hamiltonian (14) in the main text. To begin with, we should mention that our following derivations are mainly based on Ref. [46].

### 1. Effective Hamiltonian

Consider a physical system, which is described by the interaction Hamiltonian taking the form

$$H_I(t) = H_1 + H_2(t), \quad (A1)$$

with

$$H_1 = \lambda_1 \hat{O}_1 + \lambda_2 \hat{O}_2, \quad (A2)$$

$$H_2(t) = \sum_{n=1}^{N} [G_n \hat{h}_n \exp(-i\omega_n t) + G_n \hat{h}_n^\dagger \exp(i\omega_n t)], \quad (A3)$$

where $\lambda_1, \lambda_2, G_n$ are real numbers, $\hat{O}_1, \hat{O}_2, \hat{h}_n$ are the time-independent operators of the system, $N$ is the total number of different harmonic terms making up the interaction Hamiltonian, and $\omega_n > 0$. Note that in our discussion below, we assume $\omega_n \gg G_n$ (i.e., the large-detuning condition).

According to [46], the effective Hamiltonian of the system can be written as

$$H_{\text{eff}}(t) = \overline{H_I(t)} + \tfrac{1}{2}\left(\overline{[H_I(t), U_I(t)]} - [\overline{H_I(t)}, \overline{U_I(t)}]\right), \quad (A4)$$

where $\overline{H_I(t)}$, $\overline{U_I(t)}$, and $\overline{[H_I(t), U_I(t)]}$ are the time-averaged operators [46], and $U_I(t)$ is the time-evolution operator, which is given by (hereafter assuming $\hbar = 1$)

$$U_I(t) = \frac{1}{i}\int_{t_0}^{t} H_I(t')dt'. \quad (A5)$$

Based on Eqs. (A1)–(A3), it follows from Eq. (A5)

$$U_I(t) = V_I(t) - V_I(t_0), \quad (A6)$$

with

$$V_\text{I}(t) = \frac{1}{i}H_1 t + V_\text{I}'(t). \tag{A7}$$

Here,

$$V_\text{I}'(t) = \sum_{n=1}^{N} \frac{1}{\omega_n}[G_n h_n \exp(-i\omega_n t) - G_n h_n^\dagger \exp(i\omega_n t)]. \tag{A8}$$

The expression of $V_\text{I}(t_0)$ takes the same form as $V_\text{I}(t)$ with $t$ replaced by $t_0$. Inserting Eq. (A6) into Eq. (A4), we obtain

$$\begin{aligned}H_\text{eff}(t) &= \overline{H_\text{I}(t)} + \tfrac{1}{2}(\overline{[H_\text{I}(t), V_\text{I}(t) - V_\text{I}(t_0)]} \\ &\quad - [\overline{H_\text{I}(t)}, \overline{V_\text{I}(t) - V_\text{I}(t_0)}]) \\ &= \overline{H_\text{I}(t)} + \tfrac{1}{2}(\overline{[H_\text{I}(t), V_\text{I}(t)]} - \overline{[H_\text{I}(t), V_\text{I}(t_0)]} \\ &\quad - [\overline{H_\text{I}(t)}, \overline{V_\text{I}(t)}] + [\overline{H_\text{I}(t)}, V_\text{I}(t_0)]) \\ &= \overline{H_\text{I}(t)} + \tfrac{1}{2}(\overline{[H_\text{I}(t), V_\text{I}(t)]} - [\overline{H_\text{I}(t)}, \overline{V_\text{I}(t)}]), \tag{A9}\end{aligned}$$

where we have used $\overline{V_\text{I}(t_0)} = V_\text{I}(t_0)$. Here, $\overline{V_\text{I}(t)}$ and $\overline{V_\text{I}(t_0)}$ are the time-averaged operators.

Let us now give a discussion on the $\overline{H_\text{I}(t)}$, $\overline{V_\text{I}(t)}$, and $\overline{[H_\text{I}(t), V_\text{I}(t)]}$. First, according to Eqs. (A1) and (A3), we have

$$\begin{aligned}\overline{H_\text{I}(t)} &= \overline{H_1} + \overline{H_2(t)} \\ &= H_1 + \sum_{n=1}^{N}[G_n h_n \overline{\exp(-i\omega_n t)} + G_n h_n^\dagger \overline{\exp(i\omega_n t)}] \\ &= H_1, \tag{A10}\end{aligned}$$

where $\overline{\exp(-i\omega_n t)}$ and $\overline{\exp(i\omega_n t)}$ are the time-averaged quantities. Note that in Eq. (A10), we have applied $\overline{H_1} = H_1$ because $H_1$ is a time-independent operator. In addition, because of $\omega_n \gg G_n$, both $\exp(-i\omega_n t)$ and $\exp(i\omega_n t)$ are high-frequency oscillating terms and thus one can apply the rotating-wave approximation (RWA) such that [46]

$$\overline{\exp(-i\omega_n t)} = \overline{\exp(i\omega_n t)} = 0, \tag{A11}$$

which has been applied in Eq. (A10). Second, according to Eq. (A8), we have

$$\begin{aligned}\overline{V_\text{I}'(t)} &= \sum_{n=1}^{N}\frac{1}{\omega_n}[G_n h_n \overline{\exp(-i\omega_n t)} - G_n h_n^\dagger \overline{\exp(i\omega_n t)}] \\ &= 0. \tag{A12}\end{aligned}$$

Based on Eqs. (A7) and (A12), we have

$$\overline{V_\text{I}(t)} = \frac{1}{i}H_1 t + \overline{V_\text{I}'(t)} = \frac{1}{i}H_1 t. \tag{A13}$$

Therefore, according to Eqs. (A10) and (A13), we obtain

$$\begin{aligned}[\overline{H_\text{I}(t)}, \overline{V_\text{I}(t)}] &= \overline{H_\text{I}(t)}\,\overline{V_\text{I}(t)} - \overline{V_\text{I}(t)}\,\overline{H_\text{I}(t)} \\ &= \frac{1}{i}(H_1 H_1 t - H_1 H_1 t) = 0. \tag{A14}\end{aligned}$$

Next, according to Eqs. (A1) and (A7), we have

$$\begin{aligned}\overline{[H_\text{I}(t), V_\text{I}(t)]} &= \overline{H_\text{I}(t) V_\text{I}(t)} - \overline{V_\text{I}(t) H_\text{I}(t)} \\ &= \overline{[H_1 + H_2(t)][H_1 t/i + V_\text{I}'(t)]} \\ &\quad - \overline{[H_1 t/i + V_\text{I}'(t)][H_1 + H_2(t)]}\end{aligned}$$

$$\begin{aligned}&= \frac{1}{i}\overline{H_1 H_1 t} + \overline{H_1 V_\text{I}'(t)} + \frac{1}{i}\overline{H_2(t) H_1 t} \\ &\quad + \overline{H_2(t) V_\text{I}'(t)} - \frac{1}{i}\overline{H_1 H_1 t} - \frac{1}{i}\overline{H_1 t H_2(t)} \\ &\quad - \overline{V_\text{I}'(t) H_1} - \overline{V_\text{I}'(t) H_2(t)}. \tag{A15}\end{aligned}$$

Equation (A15) can be simplified as

$$\begin{aligned}\overline{[H_\text{I}(t), V_\text{I}(t)]} &= \overline{H_1 V_\text{I}'(t)} - \overline{V_\text{I}'(t) H_1} \\ &\quad + \frac{1}{i}[\overline{H_2(t) H_1 t} - \overline{H_1 t H_2(t)}] \\ &\quad + \overline{H_2(t) V_\text{I}'(t)} - \overline{V_\text{I}'(t) H_2(t)}. \tag{A16}\end{aligned}$$

Note that the Hamiltonian $H_1$ is time independent and one has $\overline{V_\text{I}'(t)} = 0$ according to Eq. (A12). Thus, we obtain

$$\begin{aligned}\overline{H_1 V_\text{I}'(t)} &= H_1 \overline{V_\text{I}'(t)} = 0, \\ \overline{V_\text{I}'(t) H_1} &= \overline{V_\text{I}'(t)} H_1 = 0. \tag{A17}\end{aligned}$$

In addition, because of Eqs. (A2) and (A3), we have

$$\begin{aligned}\overline{H_2(t) H_1 t} &= \sum_{n=1}^{N}[G_n \hat{h}_n \overline{\exp(-i\omega_n t)t} \\ &\quad + G_n \hat{h}_n^\dagger \overline{\exp(i\omega_n t)t}](\lambda_1 \hat{O}_1 + \lambda_2 \hat{O}_2), \\ \overline{H_1 t H_2(t)} &= (\lambda_1 \hat{O}_1 + \lambda_2 \hat{O}_2)\sum_{n=1}^{N}[G_n \hat{h}_n \overline{\exp(-i\omega_n t)t} \\ &\quad + G_n \hat{h}_n^\dagger \overline{\exp(i\omega_n t)t}]. \tag{A18}\end{aligned}$$

Here, it is noted that the operators $\hat{h}_n$ and $\hat{h}_n^\dagger$ may not commute with the operators $\hat{O}_1$ and $\hat{O}_2$. In the case of $\omega_n \gg \lambda_1, \lambda_2, G_n$, both $\exp(-i\omega_n t)t$ and $\exp(i\omega_n t)t$ are high-frequency oscillating terms and thus we can apply RWA such that [46]

$$\overline{\exp(-i\omega_n t)t} = \overline{\exp(i\omega_n t)t} = 0. \tag{A19}$$

Therefore, Eq. (A18) reduces to

$$\overline{H_2(t) H_1 t} = 0, \quad \overline{H_1 t H_2(t)} = 0. \tag{A20}$$

Hence, according to Eqs. (A17) and Eq. (A20), we obtain from Eq. (A16)

$$\overline{[H_\text{I}(t), V_\text{I}(t)]} = \overline{H_2(t) V_\text{I}'(t)} - \overline{V_\text{I}'(t) H_2(t)} = \overline{[H_2(t), V_\text{I}'(t)]}. \tag{A21}$$

Last, based on Eqs. (A10), (A14), and (A21), it follows from Eq. (A9)

$$H_\text{eff}(t) = H_1 + \tfrac{1}{2}\overline{[H_2(t), V_\text{I}'(t)]}. \tag{A22}$$

One can see that compared to Eq. (A9), the effective Hamiltonian (A22) has a relatively simple form. In the following, we will give a derivation on the final expression of the effective Hamiltonian $H_\text{eff}(t)$.

According to Eqs. (A3) and (A8), we have

$$\begin{aligned}\overline{[H_2(t), V_\text{I}'(t)]} &= \overline{H_2(t) V_\text{I}'(t)} - \overline{V_\text{I}'(t) H_2(t)} \\ &= \sum_{m=1}^{N}\sum_{n=1}^{N}\frac{G_m G_n}{\omega_n}\hat{h}_m h_n \overline{\exp[-i(\omega_m + \omega_n)t]}\end{aligned}$$

$$-\sum_{m=1}^{N}\sum_{n=1}^{N}\frac{G_m G_n}{\omega_n}\hat{h}_m\hat{h}_n^\dagger\overline{\exp[-i(\omega_m-\omega_n)t]}$$

$$+\sum_{m=1}^{N}\sum_{n=1}^{N}\frac{G_m G_n}{\omega_n}\hat{h}_m^\dagger h_n\overline{\exp[i(\omega_m-\omega_n)t]}$$

$$-\sum_{m=1}^{N}\sum_{n=1}^{N}\frac{G_m G_n}{\omega_n}\hat{h}_m^\dagger h_n^+\overline{\exp[i(\omega_m+\omega_n)t]}$$

$$-\sum_{n=1}^{N}\sum_{m=1}^{N}\frac{G_n G_m}{\omega_n}\hat{h}_n h_m\overline{\exp[-i(\omega_n+\omega_m)t]}$$

$$-\sum_{n=1}^{N}\sum_{m=1}^{N}\frac{G_n G_m}{\omega_n}\hat{h}_n h_m^\dagger\overline{\exp[-i(\omega_n-\omega_m)t]}$$

$$+\sum_{n=1}^{N}\sum_{m=1}^{N}\frac{G_n G_m}{\omega_n}\hat{h}_n^\dagger h_m\overline{\exp[i(\omega_n-\omega_m)t]}$$

$$+\sum_{n=1}^{N}\sum_{m=1}^{N}\frac{G_n G_m}{\omega_n}\hat{h}_n^\dagger h_m^\dagger\overline{\exp[i(\omega_n+\omega_m)t]}. \quad (A23)$$

In the case of $\omega_n \gg G_n$ and $\omega_m \gg G_m$ (large-detuning conditions), both $\exp[i(\omega_m+\omega_n)t]$ and $\exp[-i(\omega_m+\omega_n)t]$ are high-frequency oscillating terms, thus we can apply the RWA such that [46]

$$\overline{\exp[\pm i(\omega_m+\omega_n)t]} = 0. \quad (A24)$$

Assume that the inequality $|\omega_m-\omega_n| \gg \{G_m, G_n\}$ does not hold. Under this assumption, both $\exp[i(\omega_m-\omega_n)t]$ and $\exp[-i(\omega_m-\omega_n)t]$ are not high-frequency oscillating terms. In this sense, the RWA does not apply. Therefore, we have [46]

$$\overline{\exp[\pm i(\omega_m-\omega_n)t]} = \exp[\pm i(\omega_m-\omega_n)t]. \quad (A25)$$

Hence, Eq. (A23) reduces to

$$\overline{[H_2(t), V_I'(t)]} = -\sum_{m=1}^{N}\sum_{n=1}^{N}\frac{G_m G_n}{\omega_n}\hat{h}_m h_n^\dagger \exp[-i(\omega_m-\omega_n)t]$$

$$+\sum_{m=1}^{N}\sum_{n=1}^{N}\frac{G_m G_n}{\omega_n}\hat{h}_m^\dagger h_n \exp[i(\omega_m-\omega_n)t]$$

$$-\sum_{n=1}^{N}\sum_{m=1}^{N}\frac{G_n G_m}{\omega_n}\hat{h}_n h_m^\dagger \exp[-i(\omega_n-\omega_m)t]$$

$$+\sum_{n=1}^{N}\sum_{m=1}^{N}\frac{G_n G_m}{\omega_n}\hat{h}_n^\dagger h_m \exp[i(\omega_n-\omega_m)t]. \quad (A26)$$

It is straightforward to see that Eq. (A26) can be formulated as [46]

$$\overline{[H_2(t), V_I'(t)]} = 2\sum_{m=1}^{N}\sum_{n=1}^{N}\frac{G_m G_n}{\overline{\omega}_{mn}}[\hat{h}_m^\dagger, h_n]\exp[i(\omega_m-\omega_n)t], \quad (A27)$$

where

$$\frac{1}{\overline{\omega}_{mn}} = \frac{1}{2}\left(\frac{1}{\omega_m}+\frac{1}{\omega_n}\right). \quad (A28)$$

Substituting Eq. (A27) into (A22), it follows from Eq. (A22)

$$H_{\text{eff}}(t) = H_1 + \sum_{m=1}^{N}\sum_{n=1}^{N}\frac{G_m G_n}{\overline{\omega}_{mn}}[\hat{h}_m^\dagger, h_n]\exp[i(\omega_m-\omega_n)t], \quad (A29)$$

which is the final expression of the effective Hamiltonian derived from the original Hamiltonian $H_I(t)$ in Eq. (A1).

Note that in the above derivation, we have applied the conditions

$$\omega_n \gg \lambda_1, \lambda_2, G_n, \quad (A30)$$

in order to eliminate the high-frequency oscillating terms. In the following, based on the effective Hamiltonian (A29), we will derive the Hamiltonians (13) and (14) in the main text.

### 2. Derivation of the Hamiltonian (13)

Let us first show how to obtain the Hamiltonian (13) in the main text, by starting with the Hamiltonian (12) there. In this case, we have $H_I(t) = H_{I,1}$, where $H_I(t)$ is the above Hamiltonian in Eq. (A1) while $H_{I,1}$ is the Hamiltonian (12) in the main text. According to Eq. (12) in the main text, we have

$$H_I(t) = H_{I,1} = g_1(e^{i\Delta_1 t}\hat{a}_1^\dagger \sigma_{fg}^- + e^{-i\Delta_1 t}\hat{a}_1\sigma_{fg}^+)$$
$$+ g_2(e^{i\Delta_2 t}\hat{a}_2^\dagger \sigma_{fe}^- + e^{-i\Delta_2 t}\hat{a}_2\sigma_{fe}^+). \quad (A31)$$

Comparing Eq. (A31) with (A1)–(A3), we have

$$H_1 = 0, \quad N = 2, \quad (A32)$$

$$G_1 = g_1, \quad G_2 = g_2, \quad \omega_1 = \Delta_1, \quad \omega_2 = \Delta_2, \quad (A33)$$

and

$$\hat{h}_1^\dagger = \hat{a}_1^\dagger \sigma_{fg}^-, \quad \hat{h}_1 = \hat{a}_1 \sigma_{fg}^+,$$
$$\hat{h}_2^\dagger = \hat{a}_2^\dagger \sigma_{fe}^-, \quad \hat{h}_2 = \hat{a}_2 \sigma_{fe}^+. \quad (A34)$$

In the case when $\omega_n \gg G_n$ ($n = 1, 2$), i.e., $\Delta_1 \gg g_1, \Delta_2 \gg g_2$, and because of Eq. (A32), i.e., $H_1 = 0, N = 2$, it follows from Eq. (A29)

$$H_{\text{eff}}(t) = \sum_{m=1}^{2}\sum_{n=1}^{2}\frac{G_m G_n}{\overline{\omega}_{mn}}[\hat{h}_m^\dagger, h_n]\exp[i(\omega_m-\omega_n)t]$$

$$= \frac{G_1 G_1}{\overline{\omega}_{11}}[\hat{h}_1^\dagger, h_1] + \frac{G_2 G_2}{\hbar\overline{\omega}_{22}}[\hat{h}_2^\dagger, h_2]$$

$$+ \frac{G_1 G_2}{\overline{\omega}_{12}}[\hat{h}_1^\dagger, h_2]\exp[i(\omega_1-\omega_2)t]$$

$$+ \frac{G_2 G_1}{\overline{\omega}_{21}}[\hat{h}_2^\dagger, h_1]\exp[i(\omega_2-\omega_1)t]. \quad (A35)$$

Based on Eqs. (A28) and (A33), the Hamiltonian (A35) becomes

$$H_{\text{eff}}(t) = \lambda_1 [\hat{h}_1^\dagger, h_1] + \lambda_2 [\hat{h}_2^\dagger, h_2] + \lambda [\hat{h}_1^\dagger, h_2] \exp[i\delta t]$$
$$+ \lambda [\hat{h}_2^\dagger, h_1] \exp[-i\delta t], \quad (A36)$$

where $\lambda_1 = g_1^2/\Delta_1$, $\lambda_2 = g_2^2/\Delta_2$, $\lambda = (g_1 g_2/2)(1/\Delta_1 + 1/\Delta_2)$, and $\delta = \Delta_1 - \Delta_2$. In the following, we assume $\delta > 0$ which can be made by setting $\Delta_1 > \Delta_2$.

According to Eq. (A34), a simple calculation gives

$$[h_1^\dagger, h_1] = \hat{a}_1^\dagger \hat{a}_1 \sigma_{gg} - \hat{a}_1 \hat{a}_1^\dagger \sigma_{ff},$$
$$[h_2^\dagger, h_2] = \hat{a}_2^\dagger \hat{a}_2 \sigma_{ee} - \hat{a}_2 \hat{a}_2^\dagger \sigma_{ff},$$
$$[h_1^\dagger, h_2] = \hat{a}_1^\dagger \hat{a}_2 \sigma_{eg}^-,$$
$$[h_2^\dagger, h_1] = \hat{a}_1 \hat{a}_2^\dagger \sigma_{eg}^+. \quad (A37)$$

After placing Eq. (A37) in (A36), we thus obtain from Eq. (A36)

$$H_{\text{eff}}(t) = \lambda_1 (\hat{a}_1^\dagger \hat{a}_1 \sigma_{gg} - \hat{a}_1 \hat{a}_1^\dagger \sigma_{ff}) + \lambda_2 (\hat{a}_2^\dagger \hat{a}_2 \sigma_{ee} - \hat{a}_2 \hat{a}_2^\dagger \sigma_{ff})$$
$$+ \lambda \hat{a}_1^\dagger \hat{a}_2 \sigma_{eg}^- \exp(i\delta t) + \lambda \hat{a}_1 \hat{a}_2^\dagger \sigma_{eg}^+ \exp(-i\delta t), \quad (A38)$$

which is exactly the Hamiltonian $H_{e,1}$ of Eq. (13) in the main text.

### 3. Derivation of the Hamiltonian (14)

Let us now show how to obtain the Hamiltonian (14) in the main text, by starting with the Hamiltonian (13) there. Note that the Hamiltonian (13) in the main text is the same as the Hamiltonian (A38). In this sense, the above Hamiltonian $H_I(t)$ in Eq. (A1) is the Hamiltonian (13) in the main text or the Hamiltonian (A38).

By comparing Eq. (A38) with (A1)–(A3), one can find

$$H_1 = \lambda_1 (\hat{a}_1^\dagger \hat{a}_1 \sigma_{gg} - \hat{a}_1 \hat{a}_1^\dagger \sigma_{ff}) + \lambda_2 (\hat{a}_2^\dagger \hat{a}_2 \sigma_{ee} - \hat{a}_2 \hat{a}_2^\dagger \sigma_{ff}),$$
$$G_1 = \lambda, \quad \omega_1 = \delta, \quad N = 1, \quad (A39)$$

and

$$\hat{h}_1^\dagger = \hat{a}_1^\dagger \hat{a}_2 \sigma_{eg}^-, \quad \hat{h}_1 = \hat{a}_1 \hat{a}_2^\dagger \sigma_{eg}^+. \quad (A40)$$

In the case when $\omega_n \gg \lambda_1, \lambda_2, G_n$ ($n = 1$), i.e., $\delta \gg \{\lambda_1, \lambda_2, \lambda\}$, and because of $N = 1$, it follows from Eq. (A29)

$$H_{\text{eff}}(t) = H_1 + \sum_{m=1}^{1} \sum_{n=1}^{1} \frac{G_m G_n}{\bar{\omega}_{mn}} [\hat{h}_m^\dagger, h_n] \exp[i(\omega_m - \omega_n)t]$$
$$= H_1 + \frac{G_1^2}{\bar{\omega}_{11}} [\hat{h}_1^\dagger, h_1]. \quad (A41)$$

Based on Eqs. (A28) and (A39), the Hamiltonian (A41) becomes

$$H_{\text{eff}}(t) = \lambda_1 (\hat{a}_1^\dagger \hat{a}_1 \sigma_{gg} - \hat{a}_1 \hat{a}_1^\dagger \sigma_{ff}) + \lambda_2 (\hat{a}_2^\dagger \hat{a}_2 \sigma_{ee} - \hat{a}_2 \hat{a}_2^\dagger \sigma_{ff})$$
$$+ \chi [\hat{h}_1^\dagger, h_1], \quad (A42)$$

where $\chi = \lambda^2/\delta$.

According to Eq. (A40), a simple calculation gives

$$[h_1^\dagger, h_1] = \hat{a}_1^\dagger \hat{a}_1 \hat{a}_2 \hat{a}_2^\dagger \sigma_{gg} - \hat{a}_1 \hat{a}_1^\dagger \hat{a}_2^\dagger \hat{a}_2 \sigma_{ee}. \quad (A43)$$

After inserting Eqs. (A43) into (A42), we obtain

$$H_{\text{eff}}(t) = \lambda_1 (\hat{a}_1^\dagger \hat{a}_1 \sigma_{gg} - \hat{a}_1 \hat{a}_1^\dagger \sigma_{ff}) + \lambda_2 (\hat{a}_2^\dagger \hat{a}_2 \sigma_{ee} - \hat{a}_2 \hat{a}_2^\dagger \sigma_{ff})$$
$$+ \chi (\hat{a}_1^\dagger \hat{a}_1 \hat{a}_2 \hat{a}_2^\dagger \sigma_{gg} - \hat{a}_1 \hat{a}_1^\dagger \hat{a}_2^\dagger \hat{a}_2 \sigma_{ee}), \quad (A44)$$

which is exactly the Hamiltonian $H_{e,1}$ of Eq. (14) in the main text.

---